\documentclass[useAMS,usenatbib]{mn2e}

\usepackage{graphicx}
\def\LM{Lieu \& Mittaz~}
\def\LMs{Lieu \& Mittaz}

\title[The effect of lensing on the CMB]{The Effect of Lensing on the Large-Scale CMB Anisotropy}
\author[T. Shanks]{T. Shanks\thanks{E-mail: tom.shanks@durham.ac.uk}\\
Department of Physics, Durham University, South Road, Durham DH1 3LE, UK}
\begin{document}

\date{Received ; in original form 2006 September}

\pagerange{\pageref{firstpage}--\pageref{lastpage}} \pubyear{2006}

\maketitle

\label{firstpage}

\begin{abstract}
We first compare the CMB lensing model of Seljak (1996) with the
empirical model of \LM (2005) to determine if the latter approach
implies a larger effect on the CMB power-spectrum. We find that the
empirical model gives significantly higher results for the magnification
dispersion, $\sigma$, at small scales ($\theta<30'$) than that of Seljak (1996),
assuming standard cosmological parameters. However, when the empirical
foreground model of \LM is modelled via correlation functions  and used
in the Seljak formalism, the agreement is considerably improved at small
scales. Thus we conclude that the main difference between these results
may be  in the different  assumed foreground mass distributions. We
then show that a foreground mass clustering with $\xi(r)\propto r^{-3}$
gives an rms lensing magnification which is approximately constant with
angle, $\theta$. In Seljak's formalism, we show this can lead to a
smoothing of the CMB power-spectrum which is proportional to 
$e^{-\sigma^2 l^2/2}$ and which may be able to move the first CMB
acoustic peak to smaller $l$, if the mass clustering amplitude is high
enough. Evidence for a high amplitude of mass clustering comes from the
QSO magnification results of Myers et al (2003, 2005) who suggest that
foreground galaxy groups may also be more massive than expected,
implying that $\Omega_m\approx1$ and that there is strong galaxy
anti-bias, $b\approx0.2$. We combine the above results to demonstrate
the potential effect of lensing on the CMB power spectrum by showing
that an inflationary  model with neither CDM nor a cosmological constant
and that predicts a primordial first peak at $l=330$ might then fit at
least the first acoustic peak of the WMAP data. This model may be
regarded as somewhat contrived since the fit also requires high redshift
reionisation at the upper limit of what is allowed by the WMAP
polarisation results. But given the finely-tuned nature of
the standard $\Lambda$CDM model, the contrivance may  be small in
comparison and certainly the effect of lensing and other foregrounds may
still have a considerable influence on the cosmological interpretation
of the CMB.

\end{abstract}


\section{Introduction}

The standard $\Lambda$CDM model gives an excellent fit to the CMB power
spectrum from WMAP and other experiments (Hinshaw et al. 2003, 2006).
The reduced chi-squared values are impressive for a model with 7 basic
parameters and fitting over 40 approximately independent data points.
The model also appears to give an excellent fit to independent datasets
such as the 2dFGRS galaxy power spectrum (Cole et al. 2005).

Impressive though these fits are, the standard model still appears to
have a variety of more fundamental problems (see Shanks, 2005 and
references therein). For example, the CDM particle still has its own
associated problems. The limits from the Cryogenic Dark Matter Search (Akerib et al, 2004) and
other experimental constraints when combined with the WMAP result for
the CDM particle density, now leave only a small amount of `generic'
parameter space allowed by some of the basic supersymmetric models for the
neutralino (e.g. Ellis et al 2003, Barenboim \& Lykken 2006); this is
known as the `neutralino overdensity' problem. There is also what is
known as the `gravitino problem'. If the neutralino is the lightest supersymmetric particle  and the
gravitino is the next-to-lightest then if the neutralino is stable as required for a
CDM candidate then the gravitino may be also massive and unstable; if
its lifetime is longer than 0.01 secs then the decay of these particles
into other particles such as baryons could upset the Big Bang prediction
of the light element abundances (e.g. Cyburt et al 2003). Thus the usual
agreement of light element abundances with the baryon density may not be
taken for granted if the neutralino is the CDM particle. In some sense,
the invoking of the CDM particle could appear to have over-specified the
model in terms of solving the baryon nucleosynthesis problem.

The cosmological constant or dark energy presents further fundamental
problems. Indeed, string theory prefers a negative $\Lambda$ rather than
the observed positive $\Lambda$ (e.g. Witten, 2001). But it is the small
size of the cosmological constant that undoubtedly presents the biggest
difficulty for anyone assessing how successful the standard model is in
explaining the current data. In the standard model, just after inflation
the ratio of the energy density in radiation to the energy density in
the vacuum was one part in $10^{120}$. This means that the standard
model ends up potentially having more fine-tuning than implied in the
original `flatness' problem before it was solved by inflation (Guth
1981). It is this problem that presents the difficulty in assessing the
success of the standard model where the low reduced chi-squared, 
from  fits to the CMB power-spectrum has to be balanced against the
huge degree of freedom allowed by the invocation of a very small
cosmological constant.

Here we shall take the view that the fundamental complications of the
present standard model  suggest that it may still be worthwhile to
explore if other models may explain the data. This may imply the
introduction of further model parameters which may locally increase the
model's complication, but if they then allow the dropping of the
hypothesis of the cosmological constant and even the CDM particle, the
model's complication will be globally decreased.

CMB foregrounds will form our main escape route from the tight
constraints set by the CMB power-spectrum. One example will be the
evidence for the effects of high redshift ($z\approx10$) reionisation
from the WMAP polarisation results (Kogut et al 2003, Page et al 2006).
This already decreases the height of the acoustic peaks by $\approx20$\%
relative to the multipoles at larger scales even if the reionisation is
homogeneous. But galaxy clusters in  the CMB foreground at lower
redshifts provide several other ways of contaminating the CMB signal.
The hot gas in the richer clusters scatter the CMB photons via the SZ
effect and there are claims from the cross-correlation analysis of  WMAP
data with Abell clusters that the SZ decrements may extend up to 0.5
degree scales (Myers et al 2004). They can also gravitationally affect
the CMB photons primarily via the Sachs-Wolfe effect and also by
gravitationally lensing the CMB photons. Although the effects of
foreground lensing are generally claimed to be small, we shall start by
considering the claims by \LM (2005) that the effects of lensing on the CMB
power spectra may be bigger than previously estimated. \LM 
analysed lensing by foreground non-linear structures by using catalogued
properties of clusters and groups rather than power-spectra, modelling the clusters
as singular isothermal speheres.

We shall then relate the above approach to the CMB lensing formalism of
Seljak (1996) as implemented in CMBFAST (Seljak \& Zaldarriaga 1996). We
shall show that the standard formalism also allows models with a first
peak at larger wavenumber, $l$, than the standard model to fit the WMAP
data, if the mass power spectrum has a strong anti-bias with respect to the
galaxy power spectrum as suggested by the QSO lensing results of Myers
et al (2003, 2005). In this way we shall show that a large range of
models may be able to reproduce the observed first peak. In particular,
we shall argue that a low H$_0$, $\Omega_{baryon}=1$ model may give an
excellent fit to the CMB TT power spectrum with no need to invoke either
a CDM particle or dark energy.

Hence, in Section 2 we review and modify the CMB lensing results of
\LMs. In Section 3 we introduce the CMBFAST lensing formalism and
compare with the results of \LMs. In Section 4 we use these results
together with cosmological parameters determined from the QSO lensing
results of Myers et al (2003, 2005) to predict a significantly increased
effect of lensing on the CMB. In Section 5 we then consider the baryonic
cosmological model that has an intrinsic CMB first peak at
$l\approx330$. We discuss how homogeneous reionisation at high redshift
and  then the effect of lensing can significantly improve agreement with
the observed first peak at $l\approx220$.  In Section 6 we discuss our
conclusions.

\section{Lensing Model of Lieu and Mittaz (2005)}

We shall first consider the lensing model of \LM (2005). These authors
have suggested that the effects of lensing by foreground groups and
clusters on the CMB can be much larger than previously calculated. This
area has been controversial in the past with some authors claiming large
effects of lensing on the CMB  (Fukushige et al., 1994, Ellis et al.,
1998) while others find much smaller effects  (eg Bartelmann \&
Schneider, 2001). The lensing effects claimed by \LM are significantly
larger than those claimed by Bartelmann \& Schneider and somewhat larger
than claimed by Seljak. This may be because \LM use strong-lensing
formulae appropriate for clusters modelled as isothermal spheres or  NFW
profiles rather than the weak-lensing approximations used by Seljak and
Bartelmann \& Schneider, although all the above authors assume the
weak-lensing Born approximation in terms of considering only small
deviations from the original photon path. Another reason may be that
they use empirically determined models for the foreground structure
which may produce larger effects than standard $\Lambda$CDM models.

\LM treat the demagnification in the empty voids and the magnification
in the clusters separately and show that in terms of the average
magnification $\langle\eta\rangle$ they cancel, so that the same `light
conservation' result applies in this inhomogeneous case as in the
homogeneous case treated e.g. by Weinberg (1976). They show that the
same result applies when strong lensing is taken into account. Here we
follow \LM and define $\eta=(\theta'-\theta)/\theta$ as the
magnification `contrast'; the effect of the scattering is to increase
the average angular size $\theta$ of the source by this fractional
amount.

The basic relations are for the expected value of $\eta$ in the case
where the source is far behind the lens from equation (18) of \LMs:-


\begin{equation}
\langle\eta\rangle= \frac{3}{2} \Omega_{{\rm groups}} 
H_0^2 \int_0^{x_f}dx\,
[1+z(x)]\frac{(x_s-x)x}{x_s},
\end{equation}

\noindent and equation (31) of \LM gives for $\delta\eta$

\begin{equation}
(\delta \eta)^2= \frac{8 \pi^3}{3} n_0 x_f^3
\left( 1 - \frac{3x_f}{2x_s}
+ \frac{3x_f^2}{5x_s^2} \right)
\sum_{i,j} p_{ij} \frac{\sigma_i^4}{c^4} f(R_j,b_{\rm min})
\end{equation}

\noindent where

\begin{equation}
f(R_j,b_{\rm min})={\rm ln} \left(\frac{R_j}{b_{{\rm min}}}\right) - \frac{8}{\pi^2}
\end{equation}

\noindent and $p_{ij}$ is the probability of finding a group with velocity dispersion
$\sigma_i$ and radius $R_j$. \footnote[1]{Note that equation (32) of \LM (2005) contains a 
typographical error; the numerical value should read $2.7\times10^{-11}$.}

In one case, they use a space density of groups which corresponds to
that observed in the group sample of Ramella et al (1999, 2002) and
assuming the other group parameters such as the velocity dispersion,
$\sigma$. This produces  $\langle\eta\rangle\approx0.099$ and
$\delta\eta\approx0.093$, assuming the groups continue with the same
comoving space density out to $z_f=1$ in a $\Omega_m$=0.27 model. \LM
note that this assumption may not be justifiable for groups because of
the effects of dynamical evolution.  For their Abell cluster model they
find $\langle\eta\rangle\approx0.099$ and $\delta\eta=0.064$. \LM argue
that their assumption of no dynamical evolution of the cluster density
may have more empirical support in the case of clusters than groups. One
point to note is that \LM  use the line-of-sight rms velocity dispersion
and the full 3-D rms velocity dispersion should be $\sqrt3$ bigger.
Although $\langle\eta\rangle$ would remain unchanged, $\delta\eta$ would
then increase by a factor of 3, implying $\delta\eta=0.28$ for the
Ramella et al groups and $\delta\eta=0.19$ for the Abell clusters.

These values for $\delta\eta$ apply in the coherent regime of scattering
where a CMB patch has a mean angular size less than $\alpha$, the size
of a group at an average redshift. At larger scales, in the incoherent
regime of scattering, the effective $\delta\eta$ reduces according to
$\delta\eta\propto1/\theta$ (see equations 33, 34 of \LMs). For the
groups and clusters  considered by \LMs, the division between these
regimes  occurs at $\approx10$arcmin. The form of $\delta\eta$
($=\sigma(\theta)/\theta$ in the terminology of Seljak, 1996) with
$\theta$ is given in Fig. 1 for the above modified group and cluster
models of \LMs. We shall compare these to the standard CMBFAST estimates
of the magnification function in the next section.

\LM note that in cases where $\delta\eta$ is large compared to
$\langle\eta\rangle$ a skewed distribution for $\eta$ may be implied due
to negative density fluctuations cutting off at $\rho=0$ (Dyer-Roeder, 1972,
empty beam). Although this may also apply to the models above and
elsewhere in this paper, here we shall simply assume a symmetric
magnification distribution which is applied using equation A7 of Seljak
(1996).

\begin{figure}
\centering
\includegraphics[angle=0,scale=0.55]{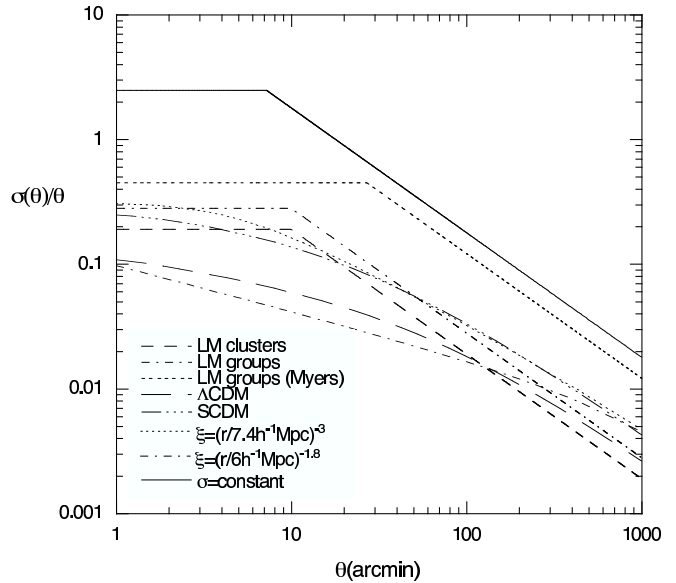}
\caption{Lensing dispersions,  $\delta\eta=\sigma/\theta$,  for the
models of \LM and Seljak (1996). The LM  clusters and groups models are
those of \LM where the lenses are respectively Abell clusters and the
groups of Ramella et al in a $\Lambda$CDM cosmology (see Sect. 2). The
results of \LM have been multiplied by $3\times$ to base the results on 
the full 3-D rms velocity dispersions rather than line-of-sight 
dispersions. The LM groups (Myers) model assumes each group contains
the high mass implied by the QSO lensing results of Myers et al. and
assumes an EdS cosmology (see Sect. 4.1). The $\Lambda$CDM and SCDM
models are the results from CMBFAST including the non-linear extensions
from Peacock \& Dodds (1994) and are comparable to the results of Seljak
(1996). The models based on the power-law correlation functions also use
Seljak's formalism; the power-law shown with index -3 assumes
$r_0=7.4$h$^{-1}$Mpc and the -1.8 power-law assumes
$r_0=6.0$h$^{-1}$Mpc. The constant $\sigma=$ (=0.005 rad.) model has an amplitude
similar to that implied by the QSO lensing results of Myers et al.
\label{fig1}}
\end{figure}

\section{Lensing Model of Seljak (1996)}

We next discuss the more usual lensing formalism of Seljak (1996) as
implemented in CMBFAST. This is based on inferring the fractional
magnification $=\sigma(\theta)/\theta$ ($=\delta\eta$ in the terminology
of \LMs) from the mass power spectrum using the variation of Limber's
formula given in equations (7) of Seljak (1996). Using the CMBFAST
source code, $\sigma(\theta)/\theta$ can be generated. The $\Lambda$CDM
model with $\Omega_m=0.27$, $\Omega_\Lambda=0.73$ (as assumed by \LMs)
with the non-linear extension of Peacock \& Dodds (1994) gives the
result shown in Fig. 1. It can be seen that the $\sigma(\theta)/\theta$
drops like $1/\theta$ at large scales as expected in the incoherent
regime while flattening at smaller scales. At $\theta>100$arcmin, the
$\Lambda$CDM model lies between the group and cluster results of \LMs.
However, at smaller scales, the models of \LM are $2-3\times$ higher
than the $\Lambda$CDM model. Tests suggest that the neglect of linear
theory evolution in the models of \LM does not appear to explain this
discrepancy. An SCDM model is also shown in Fig. 1. With COBE
normalisation, this model gives $\sigma_8=1.6$ and lies above the
$\Lambda$CDM model at all scales  and above the models of \LM at large
scales. 

The  magnification rms  dispersion, $\sigma$, is then used in Seljak's
equation A6 which, under the assumption that the isotropic term is
dominant, becomes his equation A7:

\begin{equation}
\tilde{C}(\theta)=(2 \pi)^{-1}\int_0^\infty l\,dl e^{-\sigma^2
(\theta)l^2/ 2} C_l J_0(l\theta).
\label{cthcl}
\end{equation}

\noindent where $C_l$ is the CMB angular power spectrum and
$\tilde{C}(\theta)$ is the lensed angular correlation function. This
equation has the appearance of a Hankel transform but the dependence of
$\sigma$ on $\theta$ means that this is only true under the assumption
that the $\theta$ dependence is slow.

When equation A6 is used with the standard model, only small effects are
seen on the CMB (see Seljak, Fig. 2). Indeed, we found that any
$\sigma(\theta)/\theta$  that flattens at scales smaller than  1 degree
even with amplitudes as high as $\sigma(\theta)/\theta=0.45$, generally
gave small lensing effects on the first acoustic peak. Since the
lensing dispersion is close to that of \LMs, similar results are found
under their assumptions.

A case of interest is the one where we start by assuming that
$\sigma(\theta)$ is a constant. This means that
$\sigma(\theta)/\theta\propto 1/\theta$ i.e. the usual large-scale
behaviour extends to the smallest scales. It also means that the
combination $e^{-\sigma^2(\theta)l^2/ 2} C_l$ is the lensed power
spectrum. By adjusting the size of $\sigma$ it is then possible to
smooth small scale peaks away and even to partially smooth away and
hence move the first peak. For the moment, we simply assume
$\sigma=0.005$ rad. in the case shown in Fig. 1 for $\theta>7'$. The question
is whether this amplitude and form of $\sigma(\theta)/\theta$ can be
justified physically. Equation 7 of Seljak(1996) implies that
$\sigma=$constant requires the power spectrum of the density, $P(k)$,
($\propto P_\phi\times k^4$ where $P_{\phi}$ is the power spectrum of
the potential) to have  the form $P(k)\sim k^0$ in the range of interest
which implies that $\xi(r)\sim r^{-3}$, since Fourier transforming the
density power spectrum, $P(k)\propto k^n$, to the correlation function
implies $\xi(r)\propto r^{-(3+n)}$ asymptotically for $-3<n<0$. 
Thus it appears that the constant $\sigma$ model will require a
correlation function for the mass which is steeper than the observed
galaxy correlation function, $\xi(r)\propto r^{-1.8}$. As can be seen,
the  amplitude of $\sigma$ is also much higher  than for the standard
model. But we shall see next that such an amplitude may be suggested by
considerations of QSO lensing and that the $\sigma=$constant form shown
in Fig. 1 then can be used to `move' the first acoustic peak.

We now further draw the comparison with the results of \LM by expressing
their model in power-spectrum or correlation function terms. These
authors assume a random distribution of Abell clusters each with a
singular isothermal sphere for its mass distribution. According to
Peebles (1974) a model with randomly placed monolithic clusters with
density profile $\rho(r)\propto r^{-\epsilon}$ implies a correlation
function with $\xi(r)\propto r^{-(2\epsilon-3)}$. $\epsilon=2$ as
assumed by \LM gives $\gamma=1.0$ whereas $\epsilon=2.4$ gives
$\gamma=1.8$. Fourier transforming the correlation function to the power
spectrum (as above) we find that isothermal spheres of large radius 
leads to a power spectrum of slope $k^{-2}$ which can be  assumed to
apply in the range $0.01<k<10$ h Mpc$^{-1}$.

When we take the value for the local galaxy correlation function scale
length of $r_0=6{\rm h}^{-1}$Mpc we find that, assuming $\epsilon=2.4$
for the Abell cluster model, $\sigma(\theta)/\theta$ is approximately
the same at large scales as the linear theory $\Lambda$ model (see Fig.
1). Thus, as expected, in the Seljak formalism an approximately unbiased
mass correlation function with a -1.8 power-law gives approximately the
same result for $\sigma/\theta$ as  the $\Lambda$CDM model with
non-linear extension.

If we now  assume a model in the spirit of \LM where all the mass is in
galaxies and all the galaxies are in randomly distributed Abell
clusters, this will give a mass correlation function that extends as a
-1.8 power-law to $>10$h$^{-1}$Mpc with a correlation length of
$r_0=10.6$h$^{-1}$Mpc (from equation 23 of Peebles, 1974). This then
produces  a $\sigma/\theta$ result which would be $1.7\times$ higher
than that for $r_0=6$h$^{-1}$Mpc shown in Fig. 1. This model gives a
generally flatter result with $\theta$ than the LM cluster (or group)
model, agreeing with the LM cluster model at 1 arcmin, lying a few times
lower at 10 arcmin and a few times higher at 100 arcmin. The  cut in the
cluster mass profile at radius $\approx2$h$^{-1}$Mpc assumed by \LM
makes the comparison less valid at larger scales. An $\approx2.5\times$
bigger correlation function scale length would be required to reach the
constant $\sigma$ model even at large scales ie $r_0\approx27{\rm
h}^{-1}$Mpc.  Also, as we have seen the model implied by the -1.8 power
law for $\xi$ is significantly flatter than the constant $\sigma$ model
discussed above and the SIS model with $\epsilon=2$ and $\gamma=1$ as
actually assumed by \LM would give an even flatter slope.

We conclude that \LM and Seljak appear to obtain reasonably similar
results for $\sigma/\theta$ when similar input models  are assumed.
Although the  \LM group/cluster models  show more power at small scales
than the $\Lambda$CDM model, when we assume  mass correlation functions
which should be reasonably appropriate for the \LM cluster model in the
Seljak formalism, improved agreement is seen at small scales. Therefore
in terms of the discussion (e.g. Lewis \& Challinor 2006) of how the
results of \LM relate to earlier work, there may be no serious
disagreement, with their  assumed higher  amplitude of foreground mass
clustering probably being a bigger factor in explaining any difference
than their partial  inclusion of strong lensing in their lensing
formalism. But in the wider context of affecting the first peak
position, such differences between the models are academic because none
of the $\sigma(\theta)/\theta$ models which flatten at small scales
produces the simple smoothing of the CMB first peak of the constant
$\sigma$ model; models with more small-scale power and a larger mass
clustering amplitude are required. We next look to see if there is any
evidence that the size of the magnification dispersion may be currently
underestimated.

\section{QSO magnification results}

\subsection {Group masses from QSO lensing applied to results of \LMs.}

We now consider the possible implications if galaxy group masses were as
big as those found in the QSO magnification results of Myers et al
(2003, 2005). Myers et al (2003) found that the QSO-group
cross-correlation function fitted SIS mass profiles with
$\sigma=1125$kms$^{-1}$. This velocity dispersion is comparable to the
values found for rich clusters, although the sky density of these groups
is $\approx1$deg$^{-2}$ compared to $\approx0.1$deg$^{-2}$ for the Abell
clusters. Myers et al (2005) also found that  2QZ QSO lensing by
individual galaxies again produced more anti-correlation than expected
from the standard model. Here they analysed the results using the galaxy
power spectrum after Gaztanaga (2003) and found that in the standard
cosmology an anti-bias with $b\approx0.1$ on $\approx1{\rm h}^{-1}$Mpc
scales would be required to explain the strength of anti-correlation
seen. Guimaraes, Myers \& Shanks (2005) showed that these results were
not due to selection effects by showing, for example,  that the same
cluster finding algorithms in the Hubble volume N-body simulation
produced much smaller lensing effects.

Now we discuss below other observations which disagree with the results
of Myers et al (2003, 2005). Nevertheless, here we consider the possibility of
whether galaxy groups may be dynamically young and the virialised
assumption therefore unjustified. In this case the group lensing masses
may be more correct, implying that the group masses are bigger than
expected and comparable to the mass of Abell clusters. If the full
factor of ten increase observed by Myers et al for QSO lensing over what
is expected for the concordance model is translated directly into the
results of Section 2 for CMB lensing, then $\approx3\times$ bigger
values of $\delta\eta$ will be found, resulting in
$\delta\eta\approx0.6$ for both groups and clusters in the above model.

However, Myers et al pointed out that their space density of high mass
groups implies a Universe with a mass density more consistent with that
of an Einstein-de Sitter model, rather than the $\Lambda$CDM model. The
space density of groups reported by Myers et al is
$n_0=3\pm1\times10^{-4}$h$^{3}$Mpc$^{-3}$, close to the value of Ramella
et al as used by \LMs, ($n_0=4.4\times10^{-4}$h$^{3}$Mpc$^{-3}$). Taking
a radius of 2h$^{-1}$Mpc with $b_{\rm min}=7$h$^{-1}$kpc and
substituting these in the SIS result for $\delta\eta$, we find
$\delta\eta=0.15$. Here we have assumed that the group and cluster
distribution is now unevolved out to $z_f=0.5$; our cluster and group
densities are essentially determined within this $z$ range,
reducing uncertainties due to evolution. We note that \LM find an
average value of the group velocity dispersion of 270kms$^{-1}$ whereas
the weighted sum in  their equn (32) corresponds to an average value of
460kms$^{-1}$. Clearly the more massive groups, in the tail of the
distribution, dominate the deflection effect.  If the same effect applies
for the groups of Myers et al, the rms dispersion would rise to
$\delta\eta=0.45$ in the coherent regime, as shown by the  dotted line 
marked `LM groups (Myers)' in Fig. 1.

This is the extra dispersion in the CMB anisotropy caused by coherent
scattering. Following \LM we calculate the scale above which the
scattering of light by the galaxy groups becomes incoherent rather than
coherent. In the EdS model the angular diameter distance at the average
group distance of $z=0.25$ is $d_A=500$h$^{-1}$ Mpc. Therefore our
assumed group diameter of 4h$^{-1}$Mpc subtends an angle of
$\approx0.45$deg at this redshift.  At smaller $\theta$,  $\delta\eta$
is constant, and at larger angles $\delta\eta\propto1/\theta$ (see Fig.
1). Without the effect of the tail, the results are similar to the
$\Omega_m=1$ results of Seljak (1996) although the $\delta\eta$ extends
to larger scales before starting to decrease. With the tail, the model
gives considerably higher values for $\delta\eta$ (see Fig. 1, 
dotted line) than found by Seljak for the standard $\Lambda$CDM model
(long-dashed line).

\subsection {Galaxy masses from QSO lensing applied to results of Seljak (1996).}

We next apply the galaxy lensing results of Myers et al (2005) to the
results of Seljak (1996). For the standard $\Lambda$ cosmology, Myers et
al found values of the bias $b\approx0.05-0.15$ at sub-Mpc scales,
depending on the methodology. In the EdS cosmology, they found values in
the range $b\approx0.15-0.3$. Little evidence of scale dependence was
seen in the data, so these results may be assumed to also apply at
larger scales. Scaling $\sigma/\theta$ from the unbiased results of
CMBFAST  in the $\Lambda$ and EdS cosmologies according to the above
derived anti-bias, the results for $\sigma/\theta$ clearly rise by $1/b$
i.e. by $\approx10\times$ in the $\Lambda$ case and $\approx5\times$ in
the EdS case. Clearly these results are only approximate. In a future
paper we shall relate the QSO galaxy lensing result of Myers et al
(2005) to the CMB result more directly via equations (A14) of Myers et
al (2005) and  (7) of Seljak (1996).

Although the amplitudes have risen the form of the lensing dispersions
remains the same. In particular, they flatten at small $\theta$ much
faster than the constant $\sigma$ model postulated in Section 3 above  as
having the appropriate form and amplitude required to  `move' the first
acoustic peak. The $\sigma/\theta$ result for $\xi(r)=(r/7.4h^{-1}{\rm
Mpc})^{-3}$  is shown in Fig. 1 where the amplitude has been chosen to
match the SCDM model at large scales. It can be seen that to reach the
$\sigma=$ constant model given in Fig. 1, the $r^{-3}$ mass correlation
function may even have to steepen further to ensure $\sigma/\theta$
maintains its $-1$ slope down to $7'$ scales, corresponding to
$\approx$1h$^{-1}$Mpc at an average galaxy depth of $z\approx0.15$. The
standard CDM model starts to flatten earlier at about 1 degree  and the
$\Lambda CDM$ model at yet larger scales. This means that perhaps only a
baryonic model with dissipation may produce the relatively steep  mass
correlation function in the $r<10$h$^{-1}$Mpc range of interest for
modifying the first acoustic CMB peak via lensing.

\begin{figure}
\centering
\includegraphics[angle=0,scale=0.55]{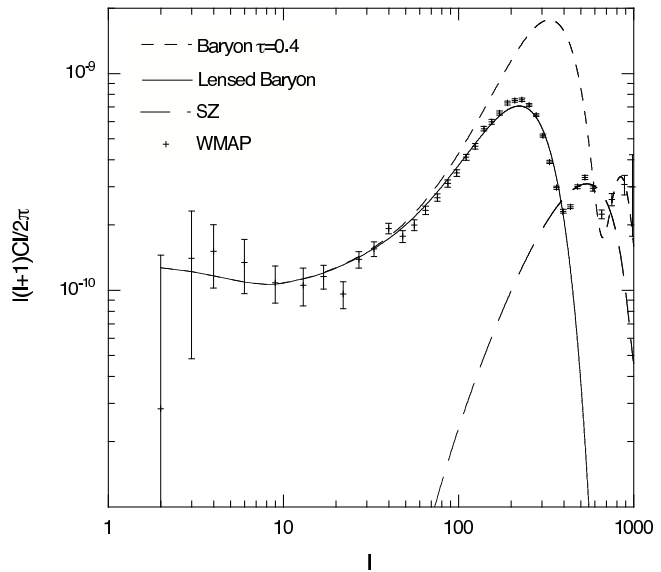} 
\caption{The $C_l$ for the $\Omega_m=1$, h=0.35 baryonic model, assuming
reionisation with optical depth, $\tau=0.4$ (dashed line). The lensed
version of the $C_l$ with $\tau=0.4$ is also shown (solid line) assuming
the constant $\sigma$ (=0.005 rad.)  model shown in Fig. 1. The
simple SZ model described in Section 5 is also shown (long-dashed line).
The data points are the 3-year WMAP results of Hinshaw et al (2006).
\label{fig2}}
\end{figure}

\section{Application to the model of Shanks (1985)}

An example of a model which initially has the first acoustic peak at
$l=330$ rather than $l=220$ is the Einstein-de Sitter, baryon dominated,
low H$_0$ model of Shanks (1985); Shanks et al. (1991); Shanks (2005). 
The $C_l$ from this model
also produces too high a peak height, so to obtain a fit to the data we
need to assume that the optical depth to reionisation is $\tau=0.4$
which is at the high end of what has been suggested by the large-scale
WMAP polarisation results (Kogut et al 2003, Page et al 2006). In fact
as shown by CMBFAST, as the $\tau$ goes up, the polarisation power
spectra quickly saturate and it becomes more difficult to reject higher
values of $\tau$. The resulting model $C_l$ is compared to the WMAP 
data in Fig. 2.

We then use the $\sigma=$ constant model from Fig. 1. There we have seen
that this model has a high amplitude but one which may be motivated by
the QSO magnification results of Myers et al (2003, 2005). The
exponentially sharp increase in smoothing as $l$ increases that then
results from equation A7 of Seljak is the  sort of behavour which might
be required to smooth the baryon model into the WMAP data. The result of
applying this lensing smoothing function is shown as the solid line in
Fig. 2 and it can be seen to give an improved fit to the first acoustic
peak. (Note that we are  assuming that $l^2C_{gl,2}(\theta)<<1$ in
equations A5 and A6 of Seljak in applying his equations A6 and A7 in
this $\sigma=$constant case where $\sigma(\theta)/\theta\approx1$ at
$\theta<20'$.) The question of whether the rise in $\sigma$ towards the
smallest scales is self-consistent with the baryon model also arises.
Certainly the linear theory baryon model does not produce such a rise,
indeed showing less small scale power than the standard model. However,
the non-linear power spectrum of the baryon model after top-down
collapse is not easy to calculate and this leaves open the possibility
that the baryon mass power spectrum may still be as steep as required to
produce the $\sigma=$ constant model. As we have seen, a mass
correlation function as steep as $r^{-3}$ approximates the behaviour
needed, compared to the observed  galaxy correlation function's
$r^{-1.8}$ behaviour. However, galaxy correlation functions as steep as
$r^{-2.2}$ are seen in samples of early-type galaxies on $r<1{\rm
h}^{-1}$Mpc scales (Zehavi et al 2005, Ross et al 2006). The amplitude
would have to be of order $r_0\approx25$h$^{-1}$Mpc implying strong anti-bias
of the type suggested by the QSO magnification results.

Note that  we have only postulated the $\sigma=$ constant model in Fig.
1 and not derived it from equation (7) of Seljak (1996).  Therefore,
although the model has $\sigma/\theta>1$ for $\theta<20$ arcmin, we have  not
violated the condition that equation (7) only applies in the case
$\sigma/\theta<1$ because of its assumption of the Born approximation.
The main assumptions we therefore have made in deriving the lensed $C_l$
in Fig. 2 is that $l^2C_{gl,2}(\theta)<<1$ in the application of
Seljak's equation A7 as noted above. Nevertheless, the lensed result in
Fig. 2 should still be regarded as only an approximation.

It has also not been possible to fit the second peak. To address this
issue, we could consider using the SZ effect caused by the known
presence of hot gas in the rich Abell clusters to re-create the second
peak after its erasure by the lensing. Myers at al (2004) have
statistically detected SZ decrements around Abell clusters in WMAP data
out to radii of 30 arcmin. These decrements simply add power to the power
spectrum on cluster-sized scales. Assuming a random distribution of
circular `clusters', with a sky density of 3deg$^{-2}$, radius 12 arcmin
and with  `decrement'  $\Delta T/T=-3\times10^{-5}$, we can produce a
quite significant `second peak' (see Fig. 1). Although the match to the
second peak is only approximate, this simple model with just a `top-hat' SZ
profile at least serves to indicate the principle of this approach. The
observed spectral index of the WMAP CMB fluctuations which effectively
constrain the SZ contribution to the first acoustic peak (Huffenberger
\& Seljak 2004) are much less effective constraints against a
significant SZ contribution to the second peak.

Further work is  needed to check if the CMB EE and TE polarisation power
spectra from WMAP can be reproduced by the above model (Kogut et al
2003, Page et al 2006). For example, it is known that lensing will
produce B modes as well as E modes in the polarisation signal
(Zaldarriaga \& Seljak 1998) and it will be interesting to see if this
provides a constraint on the current approach.

For the moment,  we conclude that using high-redshift reionisation to reduce the
first two acoustic peaks, then lensing to move the first peak to larger
scales and finally SZ to recreate the second peak, it appears possible to
achieve an approximate fit to the WMAP CMB fluctuation power spectrum.
Although this approach may be criticised as fine-tuning CMB foreground
effects to achieve a fit to the data, it may also be viewed as being much less
fine-tuned than the approach taken via exotic particles and dark energy,
as employed to construct the standard $\Lambda$CDM  model.

\section{Conclusions}

We first compared models for the effect of foreground galaxy clusters
and groups on the CMB power-spectrum. We have  shown that the lensing
effects considered by \LM may have bigger effects than they suggested,
if the full 3-D velocity dispersion is used rather than the 1-D
dispersions assumed by these authors. This change produces
$\approx3\times$ bigger results for $\delta\eta(=\sigma(\theta)/\theta)$
than they suggested, which will increase the effect on the CMB peaks.

We find that the empirical model of \LM  then gives significantly higher
results for the magnification dispersion at small scales ($\theta<30'$)
than those of Seljak (1996) assuming standard cosmological parameters.
We have also shown that under the same  foreground assumptions, now
modelled in mass correlation function terms,  the lensing formalism of
\LM produces approximately the same results  as the that of Seljak
(1996). But both formalisms in the case of the standard $\Lambda$CDM
foreground or in the empirical cases considered by \LM tend to produce
$\sigma/\theta$ relations which are too flat towards small $\theta$ and
have too small amplitudes to modify significantly the first  peak of the
CMB power spectrum.  

In conjunction with a model that assumed lensing dispersion
$\sigma(\theta)=$ constant we then used the QSO magnification results of
Myers et al (2003, 2005) to show that, if galaxy groups have bigger than
expected masses, there could be significant effects of gravitational
lensing on even the first peak of the CMB. Although the results of Myers
et al remain controversial (Scranton et al. 2005, Mountrichas \& Shanks
2007), it is clear that at least in principle, lensing can affect the
interpretation of the CMB power spectrum.  But to have a significant
effect on the first peak, the mass clustering power spectrum would then
need to have been significantly underestimated in previous studies with
anti-bias, $b\approx0.2$, being required.

There are many pieces of evidence which favour the standard model bias
value of $b\approx1$. For example, redshift space distortion analyses
imply $\beta=\Omega_m^{0.6}/b\approx0.5\pm0.1$ (e.g.  Ratcliffe et al
1996, Hawkins et al 2003.) But such analyses tend to assume that the
galaxies trace the mass with linear bias  and  in the case where groups
contain as much mass as clusters this assumption does not apply. Even
here we note that the infall velocity patterns around the rich
environments of passive galaxies in 2dFGRS are very similar to  those
around the poorer environments of active galaxies (Madgwick et al.
2003), suggestive of the evidence from the QSOs that groups dominated by
spirals may have as much associated mass as the richer environments of
the early-types.

But still, the most direct previous constraints on the mass power
spectrum come from estimates of cosmic shear. The QSO lensing results
are again in clear contradiction with these since they are mainly
consistent with $b=1$. For example, Seljak (1996, Fig. 1) suggests that
the cosmic shear upper limits on $\sigma( \theta)/\theta$  are
close to the SCDM predictions and well below the constant $\sigma$ model
in Fig. 1. Yet even in the case of cosmic shear there are possible
systematic effects which may cause underestimates of the mass density
and overestimates of $b$. For example,  Hirata \& Seljak (2004) have
suggested that intrinsic galaxy alignments may prove difficult to
disentangle from the effects of gravity on the shapes of galaxies. They
suggest that even with the availability of perfect redshift information
to remove physically associated pairs of galaxies, the effect of lensing
may still be corrupted by physical alignment if the foreground tidal
field dynamically elongates the foreground galaxy while elongating the
background galaxy via its lensing effect. This could cause the cosmic
shear to be significantly underestimated in present samples and explain
the discrepancy with the QSO lensing result (see also Mandelbaum et al., 
2006).

Although the agreement in form between the galaxy and QSO power spectra
and WMAP are impressive, there is the possibility that the galaxies do
not trace the mass on the largest scales and may be affected by
scale-dependent bias. The high masses for groups suggested by the QSO
lensing results are certainly consistent with this idea. We also must
recall that there are other measures of clustering which give much
higher amplitudes of clustering than those given by the galaxies. For
example, the rich Abell galaxy clusters have a correlation length of
$r_0\approx25$h$^{-1}$Mpc.  Another measure of clustering of mass at
large scales is our large $\approx700$kms$^{-1}$ motion with respect to
the CMB (eg Kogut et al 1993). This has always been at the upper end of
what is allowed if $\beta\approx0.5$ and these results are made even
more remarkable, given the observational evidence that the local region
of $\approx$40h$^{-1}$Mpc radius or more is also moving with a
comparable `bulk motion' (Lynden-Bell et al. 1988).

Since the non-linear evolution of CDM models is well-worked through via
collisionless N-body simulations, it is unlikely that such models can
produce anything but the standard forms for $\sigma(\theta)/\theta$.
Dissipation in  galaxy and cluster formation in the pure baryonic models
could still leave the mass power spectrum steeper than expected from
its initial, linear form.

We have taken  the view here that the crucial observation is the high
fraction of baryons in Abell clusters such as Coma. This `baryon
catastrophe' already forces the CDM model to include $\Lambda$ to allow
a model with a lower $\Omega_m$ to be compatible with zero spatial
curvature and inflation (White et al 1993). If the X-ray gas is to
explain the missing mass in Coma, routes must be found to reconcile WMAP
with the adiabatic $\Omega_{baryon}=1$ prediction and we have shown that
cosmic CMB foregrounds may form such a route. If the model can be shown
to satisfy other constraints such as the WMAP polarisation power
spectra, the result could be a model which is less finely-tuned than the
standard model and is based on known physics, without recourse to
undetected neutralinos or postulated dark energy.

\section*{Acknowledgments}

Richard Lieu is gratefully acknowledged for  useful discussions. The
referee is also thanked for comments which have significantly improved
the clarity of the paper.\\

\noindent
{\bf References}

\noindent
Akerib et al 2004, Phys. Rev. D72, 052009  

\noindent
Barenboim, G. \& Lykken, J., 2006, JHEP, 12, 5

\noindent
Bartelmann, M. \& Schneider, P., 2001, Physics Reports, 340, 291.

\noindent
Cole, S.M. et al 2005, MNRAS, 362, 505

\noindent 
Cyburt,ÊR.H., Ellis,ÊJ., Fields,ÊB.D., Olive,ÊK.A., 2003, Phys. Rev. D, 67, 103521

\noindent
Dyer, C.C. \& Roeder R.C., 1972, ApJ, 174, L115.

\noindent
Ellis, G. F. R., Bassett, B. A. C. C. \& Dunsby, P. K. S.,  1998,
Classical and Quantum Gravity, 15, 2345

\noindent
Ellis, J., Olive,ÊK.A., Santoso,ÊY. \& Spanos,ÊV.C. 2003 Physics Letters B, 565, 176.

\noindent
Fukushige, T., Makino, J., \& Ebisuzaki, T. 1994, ApJ 436, L107.

\noindent
Gaztanaga, E. 2003, ApJ, 589, 82.

\noindent
Guimaraes,ÊA.C.C., Myers,ÊA.D., Shanks,ÊT., 2005, MNRAS, 362, 657.

\noindent
Guth, A.H., 1981, Phys. Rev. D, 23, 347.

\noindent
Hawkins, E. et al. 2003, MNRAS, 346, 78.

\noindent 
Hinshaw, G. et al., 2003, ApJS, 148, 63.

\noindent 
Hinshaw, G. et al., 2006, preprint astro-ph/0603451.

\noindent
Hirata, C.M. \& Seljak, U. 2004 Phys. Rev. D, 70, 6, 063526

\noindent
Huffenberger, K.M. \& Seljak, U., 2004 Phys. Rev. D, vol. 70, 6, 063002

\noindent
Kogut, A. et al., 1993, ApJ, 419, 1.

\noindent
Kogut, A. et al., 2003, ApJS, 148, 161.

\noindent
Lewis, A. \& Challinor, A. 2006, Physics Reports, 429, 1.

\noindent
Lieu, R., \& Mittaz, J.P.D., 2005, ApJ, 628, 583 

\noindent
Lynden-Bell, D., Faber, S.M. Burstein, D., Davies, R.L., Dressler, A., Terlevich, R.J.
\& Wegner, G. 1988, ApJ, 326, 19.

\noindent
Madgwick, D.S. et al., 2003, MNRAS, 344, 847.

\noindent
Mandelbaum, R., Hirata, C.M., Ishak, M., Seljak, U., Brinkman, J., 2006 MNRAS 367, 611

\noindent
Mountrichas, G., \& Shanks, T., 2007, preprint astro-ph/0701870

\noindent 
Myers, A.D., Shanks, T.,  Boyle,ÊB.ÊJ., Croom,ÊS.ÊM., Loaring,ÊN.S., 
Miller,ÊL. \& Smith,ÊR.J. 2003, MNRAS, 342, 467.

\noindent
Myers, A.ÊD., Shanks, T., Outram, P.J., Frith, W.J. \& Wolfendale,ÊA.W.
2004, MNRAS, 347, L67.

\noindent
Myers, A.D., Shanks, T.,  Boyle,ÊB.ÊJ., Croom,ÊS.ÊM., Loaring,ÊN.S., 
Miller,ÊL. \& Smith,ÊR.J. 2005, MNRAS, 359, 741.

\noindent
Page, L. et al., 2006, astro-ph/0603450.

\noindent
Peacock, J.A. \& Dodds, S.J., 1994, MNRAS, 267, 1020.

\noindent
Peebles, P.J.E 1974 Astr. \& Astr., 32, 197.

\noindent
Ramella, M. et al, 1999, A \& A, 342, 1.

\noindent
Ramella, M., Geller, M.J., Pisani, A., \& da Costa, L.N.,
2002, AJ, 123, 2976.

\noindent
Ratcliffe, A., Shanks, T., Parker, Q.A. \& Fong, R. 1996, MNRAS, 296, 191.

\noindent
Ross, N.P. et al. 2007, astro-ph/0612400

\noindent
Scranton, R. et al 2005 ApJ, 633, 589.

\noindent
Seljak, U., 1996, ApJ, 463, 1

\noindent
Seljak, U. \& Zaldarriaga, M., 1996, ApJ, 469, 437

\noindent 
Shanks, T., 1985, Vistas in Astronomy, 28, 595.

\noindent
Shanks, T. et al., 1991, In `Observational Tests of Cosmological
Inflation', Eds. Shanks, T., Banday, A.J., Ellis, R.S., Frenk, C.S. \& Wolfendale,
A.W. pp. 205-210 Dordrecht:Kluwer.

\noindent
Shanks, T., 2005., In `Maps of the Cosmos', Proc. of IAU Symposium 216. Eds. 
M.M. Colless, L. Staveley-Smith \& R. Stathakis. San Francisco: ASP, pp. 398-408

\noindent
Weinberg, S., 1976, ApJ, 208, L1.

\noindent
White, S.D.M., Navarro, J.F., Evrard, A.E. \& Frenk, C.S, 1993 Nature 366, 429.

\noindent
Witten, E. 2001 http://theory.tifr.res.in/strings
/Proceedings/witten/22.html

\noindent
Zaldarriaga, M. \& Seljak, U. 1998, Phys.Rev.D, 58, 023003.

\noindent
Zehavi et al 2005, ApJ, 621, 22

\label {lastpage}

\end{document}